\documentclass{article}
\usepackage{spconf,amsmath,graphicx, amssymb}
\usepackage{float}
\usepackage{cite}
\usepackage{mathtools}

\usepackage{capt-of}
\usepackage{url}
\usepackage[nolist]{acronym}
\usepackage{amssymb,multirow}
\usepackage{booktabs, bm}
\usepackage{lineno}
\usepackage[font={footnotesize}]{caption}

\newacro{MVDR}[MVDR]{Minimum Variance Distortionless Response}
\newacro{GEV}[GEV]{Generalized Eigenvalue Decomposition}
\newacro{MWF}[MWF]{Multichannel Wiener Filter}
\newacro{SDW-MWF}[SDW-MWF]{Speech Distortion Weighted MWF}
\newacro{SISDR}[SI-SDR]{Scale Invariant Signal-to-Distortion Ratio}
\newacro{STFT}[STFT]{Short-Time Fourier Transform}
\newacro{DNN}[DNN]{Deep Neural Network}
\newacro{SNR}[SNR]{Signal-to-Noise-Ratio}
\newacro{FLOPs}[FLOPs]{Floating Point Operations}
\newacro{SIR}[SIR]{Signal-to-Interference Ratio}
\newacro{ASR}[ASR]{Automatic Speech Recognition}
\newacro{IBM}[IBM]{Ideal Binary Mask}
\newacro{IRM}[IRM]{Ideal Ratio Mask}
\newacro{WLM}[WLM]{Wiener-Like Mask}
\newacro{SDR}[SDR]{Signal-to-Distortion Ratio}
\newacro{STOI}[STOI]{}
\newacro{PESQ}[PESQ]{}
\newacro{WER}[WER]{Word Error Rate}

\title{Learning filterbanks for End-to-end acoustic beamforming}

\name{Samuele Cornell$^{1}$, Manuel Pariente$^2$, Fran\c{c}ois Grondin$^3$, Stefano Squartini$^{1}$}
\address{
$^1$Università Politecnica delle Marche, Italy \\
$^2$Universit\'e de Lorraine, CNRS, Inria, LORIA, France  \\
$^3$Universit\'{e} de Sherbrooke, Canada  \\
}

\begin{document}
\ninept

\maketitle

\begin{abstract}
Recent work on monaural source separation has shown that performance can be increased by using fully learned filterbanks with short windows. On the other hand it is widely known that, for conventional beamforming techniques, performance increases with long analysis windows. This applies also to most hybrid neural beamforming methods which rely on a deep neural network (DNN) to estimate the spatial covariance matrices.
In this work we try to bridge the gap between these two worlds and explore fully end-to-end hybrid neural beamforming in which, instead of using the Short-Time-Fourier Transform, also the analysis and synthesis filterbanks are learnt jointly with the DNN.
In detail, we explore two different types of learned filterbanks: fully learned and analytic. 
We perform a detailed analysis using the recent Clarity Challenge data and show that by using learnt filterbanks it is possible to surpass oracle-mask based beamforming for short windows. 
%It is widely known that for conventional beamforming techniques performance increases with long analysis windows. This applies also to hybrid neural beamforming methods which rely on a deep neural network to estimate the spatial covariance matrices.
%In this work we explore fully end-to-end 
%In this work we explore conventio
%Hybrid neural beamformers, in which deep learning techniques are used together with conventional beamforming solutions such as Minimum Variance Distortionless Response and fully neural beamformers, which instead are fully data-driven. 
%In detail, we study conventional beamforming 

%Lorem Ipsum is simply dummy text of the printing and typesetting industry. Lorem Ipsum has been the industry's standard dummy text ever since the 1500s, when an unknown printer took a galley of type and scrambled it to make a type specimen book. It has survived not only five centuries, but also the leap into electronic typesetting, remaining essentially unchanged. It was popularised in the 1960s with the release of Letraset sheets containing Lorem Ipsum passages, and more recently with desktop publishing software like Aldus PageMaker including versions of Lorem Ipsum.Lorem Ipsum is simply dummy text of the printing and typesetting industry. Lorem Ipsum has been the industry's standard dummy text ever since the 1500s, when an unknown printer took a galley of type and scrambled it to make a type specimen book. It has survived not only five centuries, but also the leap into electronic typesetting
\end{abstract}
\begin{keywords}
 acoustic beamforming, end-to-end learning, source separation, speech enhancement, multi-channel processing.  % speech recognition 
 % front-end processing 
\end{keywords}
\section{Introduction}
\label{sec:intro}
%Multi-channel front-end processing is a fundamental component in %many speech processing pipelines \cite{umbach, vincent2018audio,  %haeb2020far}. 
Most current deep learning based beamforming or \emph{neural beamforming} can be divided into two main categories: \emph{hybrid} \cite{heymann_blstm, Boeddeker_practical, xiao_mvdr, beamnet, xiao_deep, erdogan2016improved, ochiai_unified_2017, ochiai_beam_tasnet, zhang2021adl, aroudi2020dbnet, li2020deep} and \emph{fully neural} \cite{luo2019fasnet, luo2020end, luo2020implicit, xu2021generalized, li2021mimo}. Hybrid techniques couple \ac{DNN} with established beamforming methods such as \ac{MVDR} \cite{capon_mvdr}, \ac{MWF} or \ac{GEV} \cite{gev_warsitz} solutions.  
Usually they employ a \ac{DNN} to estimate the spatial covariance matrix (SCM) via a time-frequency mask \cite{heymann_blstm, Boeddeker_practical, xiao_mvdr, beamnet, xiao_deep, erdogan2016improved, ochiai_unified_2017}  in the magnitude \ac{STFT} domain.
Another approach \cite{ochiai_beam_tasnet} is to use the \ac{DNN} model to estimate the target and interferer time domain signals and subsequently derive the SCMs.
In both cases the \ac{DNN} is usually a monaural model and the mask is estimated on one microphone channel used as a reference. Additional spatial features are sometimes used to improve the masks estimation \cite{aroudi2020dbnet, zhang2021adl}. As they rely on SCM estimation to derive the beamforming solution, hybrid neural beamformers performance is greatly affected by the frame size used to estimate the SCMs of the target and interferer/noise signals.

% doa estimation methods too ? see https://hal.inria.fr/hal-02355613v2/document
% %This leads to a trade-off between performance and latency for near real-time applications \cite{luo2019fasnet} which can be only partially alleviated by using asymmetrical \ac{STFT} analysis and synthesis windows \cite{}. 
% francois here could help 

%The analysis window determines the frame size for each beamforming filter and  because for a robust estimation of the SCMs a sufficiently long context is needed. 

On the other hand, fully neural models employ a \ac{DNN} to directly estimate the beamforming filters \cite{luo2019fasnet, luo2020end} or even the time domain target signal directly \cite{luo2020implicit, qi2020exploring, fu2021desnet}. 
Being fully data-driven these methods are less sensitive to the frame size of the beamforming filters. For example FasNet \cite{luo2019fasnet} is able to reach comparable or superior performance with respect to conventional oracle beamformers for low latency applications with remarkably smaller frame size and latency. This is aligned with results in monaural source separation, where fully learned representations have been shown to surpass the STFT in both clean \cite{luo2018convtasnet} and noisy conditions \cite{pariente2020filterbank} especially for short windows \cite{luo2020dualpath}. 
However, fully neural models are arguably ``less interpretable" and are prone to introducing more non-linear distortion than conventional beamformers. FasNet \cite{luo2019fasnet} is a notable exception as it estimates linear spatial filters for filter-and-sum beamforming, thus enabling to e.g. visualize the beam-patterns. This however is not possible for other methods \cite{luo2020implicit, qi2020exploring, fu2021desnet} as the multi-channel processing is done inside the DNN. 

% This could cause %Nonetheless fully neural models are desirable due to often being 

%This could cause a degradation in performance for back-end applications such as e.g. \ac{ASR}. While end-to-end training/adaptation with the back-end model \cite{} can solve this issue, 

In this paper we attempt to bridge the gap between these two paradigms and study conventional beamforming with fully learned filterbanks; inspired by aforementioned promising results achieved in monaural source separation \cite{luo2018convtasnet, pariente2020filterbank, luo2020dualpath}.
We propose to train a hybrid neural beamformers where the \ac{DNN} is used to estimate the SCMs via a mask. Unlike previous works \cite{heymann_blstm, Boeddeker_practical, xiao_mvdr, beamnet, xiao_deep, erdogan2016improved, ochiai_unified_2017, ochiai_beam_tasnet, zhang2021adl, aroudi2020dbnet, li2020deep} we learn the analysis and synthesis filterbanks in place of the \ac{STFT} along with the mask-estimation \ac{DNN} using time-domain losses. 
We consider for this study fully unconstrained linear filterbanks as used in \cite{luo2018convtasnet}
and the recently proposed learnable analytic filterbanks \cite{pariente2020filterbank} which allow for shift invariance, an especially desirable property in this case. 

%As a parallel contribution, we place here particular attention to low latency causal beamforming. 
%In fact, most neural beamforming techniques \cite{luo2020end, luo2020implicit} with few exceptions \cite{Boeddeker_practical, luo2019fasnet} focus on offline applications where it is assumed that the whole signal is available to estimate the beamforming filters. 
%We argue however that these assumptions are hardly satisfied in most real-world applications such as speech-based virtual assistant devices, autonomous robots, hearing aids, video-conferencing et cetera, where it is required real-time or near-real time enhancement. 

%For these reasons we place particular attention to low latency causal beamforming and use the recently proposed Clarity Challenge dataset to perform our experiments. 

Our findings suggest that, even with a simple hybrid neural beamforming model, performing \ac{MVDR} and \ac{MWF} in a learned representation can achieve comparable or superior denoising performance than oracle STFT based beamformers. In detail, regarding MVDR we show that learned filterbanks are consistently able to provide better \ac{SISDR} improvement over STFT-based models and, even, oracle \ac{WLM} \cite{erdogan_psm} for both small and large analysis windows. Notably, the best learned filterbanks model outperform the best oracle \ac{WLM} configuration by more than 2\,dB. 
Instead, regarding MWF, we show that a significant performance gain over oracle \ac{WLM} can be achieved for small windows. For longer windows the oracle \ac{WLM} fares better. 
In general we found learned filterbanks models to be superior to STFT-based non-oracle ones with the ones based on analytic filterbanks performing the best. We make our code publicly available through the Asteroid Source Separation toolkit \cite{pariente2020asteroid} \footnote{github.com/asteroid-team/asteroid}. 

%Deep neural network based source separation has seen tremendous progress in recent years. This progress has been made possible both by powerful DNN architectures able to model long contexts and the adoption of fully learnable front-ends \cite{} with overcomplete basis. 
%In parallel 
%These systems can be roughly classified into \emph{fully neural} and \emph{hybrid}

\vspace{-0.2cm}
\section{Acoustic Beamforming with learned filterbanks}\label{sec:learned_beam}
Considering an array of $M$ microphones we can denote with $\bm{y}(t) = \left[y_{1}(t), y_{2}(t), \hdots, y_{M}(t)  \right]^{T}$ the matrix of the time-domain signals at each microphone, with $t$ being the sample index. 
We consider here a situation where $\bm{y}(t)$ is comprised of two terms:
\vspace{-0.2cm}
\begin{equation}
    \bm{y}(t) = \bm{x}(t) + \bm{\nu}(t), 
\end{equation}

with $\bm{x}(t)$ the matrix of desired source signals and $\bm{\nu}(t)$  the matrix of interfering source signals at the microphones. 
Our goal here is recovering the desired signal $x_{r}(t)$  at an arbitrarily chosen reference microphone $1 \leq r \leq M$ by suppressing the interferer.
This implies that, in this study, the target is a reverberated source signal and joint enhancement and dereverberation is left for future work. Accordingly, the target signal at reference microphone $r$ is given by $x_r(t) = \sum_{\tau=1}^{L_h} h_r(\tau) x^a(t-\tau)$, where $x^a(t)$ is the dry desired source signal and $h_r(\tau)$ is the impulse response of length $L_h$ characterizing the acoustic propagation of the desired source signal to the reference microphone at time lag $\tau$.
%We do not perform dereverberation here and leave it for future work. 
%This can be achieved by conventional beamforming techniques 
%training a suitable \ac{DNN}-based beamforming model depicted in Figure \ref{fig:mask_beam}.
Recovering of $x_{r}(t)$ can be achieved by conventional spatial filtering techniques if an estimate of the target signal and the interferer SCMs can be produced. 

\begin{figure}
\centering
\includegraphics[width=6.4cm]{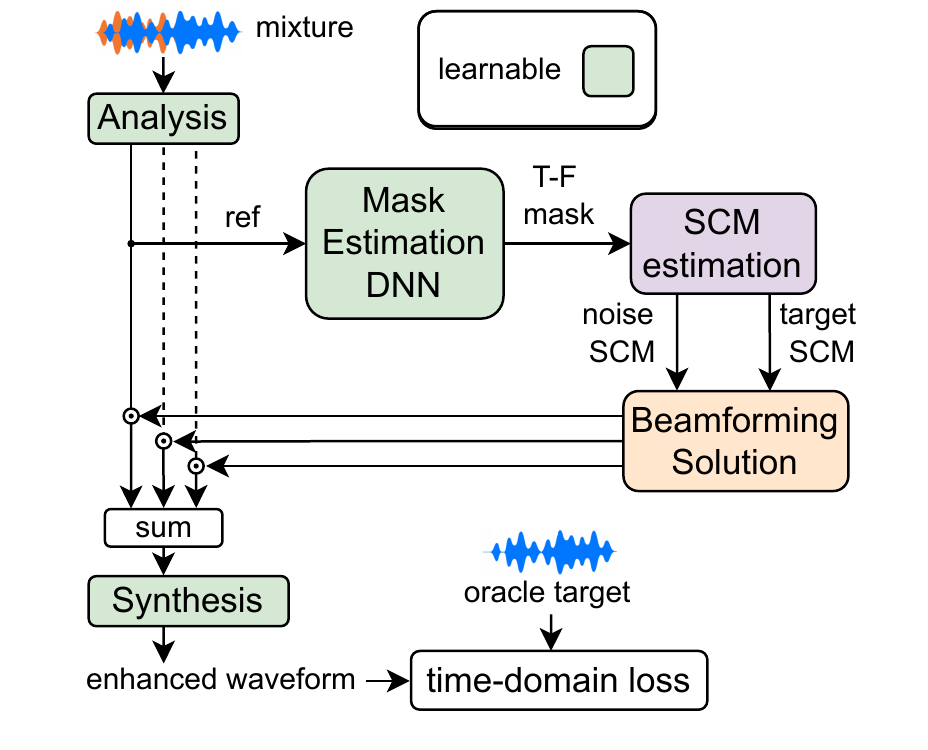}
\vspace{-0.4cm}
\caption{Framework overview. The gradient is back-propagated from waveform domain. This allows to learn the analysis and synthesis filterbanks along with the mask-estimation \ac{DNN}.}
%-based method here.}
\label{fig:mask_beam}
\vspace{-0.6cm}
\end{figure}

In this paper we follow a simple hybrid neural beamforming framework, illustrated in Figure \ref{fig:mask_beam}, where such estimates are produced by a monaural mask estimation \ac{DNN} $\mathcal{F}(\cdot, \bm{\theta})$ with $\bm{\theta}$ trainable parameters. 
An STFT analysis filterbank $\bm{\phi}_n(t)$ is used to extract the time-frequency representation for every $m$-th microphone input signal, obtaining a third order tensor:
\vspace{-0.2cm}
\begin{equation}
    Y_m (n, k) = \sum_{t=1}^{L} y_{m}(t+ kH) \phi_n(t), \,\,\, n \in \left[1, \hdots, N   \right],
\end{equation}

where $\left\{  \phi_n(t) \right\}_{n=\left[1, \hdots N\right]}$ are the $N$ STFT analysis filters each of size $L=N$ and $H$ is the hop-size or stride factor. Consequently $n$ and $k$ denote respectively the frequency bin and frame indexes. %Consequently $\bm{Y}_m (f, k) \in \mathbb{C}^{M \times N \times K}$ where K is the number of frames for the current input signal. 

The mask-estimation \ac{DNN} is given as input features this complex STFT representation (real and imaginary part)
at a chosen reference channel $r$ and outputs a mask $m(n, k)$ for the target signal in the magnitude STFT domain (i.e. with shared values between real and imaginary parts): 
\vspace{-0.2cm}
\begin{equation}
    m(n, k) = \sigma(\mathcal{F}(Y_{m=r} (n, k) , \bm{\theta}))
\end{equation}

where $\sigma(\cdot)$ denotes the sigmoid activation.
The interferer signal mask is obtained simply as $1- m(n, k)$. We found this configuration to work the best in our experiments rather than outputting two distinct masks and/or using a different activation (e.g. softmax). More in detail, this configuration led to more stable training with the dataset used in our experiments, while the use of two distinct masks often led to ill-conditioned SCMs especially when the activation was unbounded (e.g. ReLU).

These masks are then used to compute the frame-wise SCMs of target and interferer respectively:

\vspace{-0.2cm}
\begin{equation}\label{eq:scm_estimation}
\begin{aligned}
    \bm{R}_{x}(n, k) = \bm{Y} (n, k)  m(n, k)  \bm{Y} (n, k)^{H}, \\ 
    \bm{R}_{\nu}(n, k) = \bm{Y} (n, k)  ( 1- m(n, k) )  \bm{Y} (n, k)^{H},
\end{aligned}
\end{equation}
where $H$ denotes the Hermitian transpose and both target and interferer SCMs are $4$-th order tensors $\in \mathbb{C}^{M \times M \times N \times K}$, where $K$ is the number of frames.
In this study, for simplicity, we consider non-causal systems. In this instance, following previous works \cite{heymann_blstm, Boeddeker_practical, xiao_mvdr,xiao_deep, beamnet, erdogan2016improved, ochiai_unified_2017, ochiai_beam_tasnet}, the overall SCM can be computed by simply averaging the frame-wise SCM over the whole input mixture segment: $\bm{R}_{\rho}(n) = \frac{1}{K} \sum_k \bm{R}_{\rho}(n, k)$ for both target $\rho = x$ and interferer $\rho = n$.
In addition to non-causality, this averaging operation requires that the transfer functions of the target source and interferer do not change in the time-frame over which the averaging is performed (e.g. for the target, $h_r$ is assumed stationary).

%For causal systems instead cumulative mean at each $k$-th frame can be used: $\bm{R}^{\rho}_{m,m}(n, k) = \frac{1}{k} \sum_i^{k} \bm{R}^{\rho}_{m,m}(n, k), 1 \leq k \leq K$ as information of future frames is not available; note that for $k=K$ the two quantities coincide. 

From such estimated SCMs different beamforming solutions can be computed. In this study we consider \ac{MVDR} and \ac{MWF}. %which have different and opposite behaviours regarding interferer suppression and distortion. 
%Importantly, differently from \ac{GEV}, these beamforming solutions lend themself well for end-to-end training with waveform-level loss, a necessity if we wish to learn the analysis and synthesis transforms. 
%\ac{GEV} in fact does not preserve phase alignment with the reference channel  \cite{heymann_blstm, luo2019fasnet}. This makes it difficult to perform end-to-end training at waveform level in a straightforward way and requires novel solutions that are behind the scope of this work. 

Regarding MVDR, we use the formulation from \cite{souden2009optimal} and estimate the spatial filter as 
\begin{equation}
    \bm{w}_m^{MVDR}(n) = \frac{\bm{R}^{-1}_{\nu}(n) \bm{R}_{x}(n) \bm{u_{m}}}{tr\left\{ \bm{R}^{-1}_{\nu}(n)\bm{R}_{x}  (n) \right\}},
\end{equation}
where $tr\left\{ \cdot \right\}$ denotes the trace operator and $\bm{u_{m}}$ is an one-hot column vector for which the $m$-th term is $1$, and all others are $0$.
Regarding MWF instead we simply obtain the filter coefficients as:

\begin{equation}
    \bm{w}_m^{MWF}(n) = \left(\frac{1}{\bm{R}_{x}(n) + \bm{R}_{\nu}(n)}\bm{R}_{x}(n)\right) \bm{u_{m}},
\end{equation}

\noindent
and the beamformed signal is obtained as 
\begin{equation}
    \tilde{X}(n, k) = \bm{w}_m(n)^H  \bm{Y}(n, k),
\end{equation}
\noindent
which is finally brought back to time-domain via a \emph{synthesis} inverse-STFT (iSTFT) filterbank $\bm{\psi}_n(t)$ filterbank with $N$ synthesis filters $\left\{  \bm{\psi}_n(t) \right\}_{n=1, \hdots N}$ of, again, length $L=N$ each:

\begin{equation}
    \tilde{x}(t) = \sum_{k=1}^{K} \sum_{n=1}^{N} \tilde{X}(n, k) \psi_{n}(t- kH).
\end{equation}
%Such end-to-end enhancement allows to use a time-domain loss such as SISDR \cite{le2019sdr} which has been shown to be particularly effective in source separation \cite{luo2018convtasnet} and speech enhancement \cite{kolbaek2020loss}. 

% by using the oracle 
%such time-domain estimate of the target signal at reference microphone $r$. 

\subsection{Learnable Analysis and Synthesis Filterbanks}\label{sec:leanrable}

In this work we propose to replace the STFT and iSTFT filterbanks with learnable linear filterbanks and perform spatial filtering in a learned linear basis. These filterbanks are learnt end-to-end jointly along with the mask-estimation DNN $\mathcal{F}(\cdot, \bm{\theta})$ as the gradient can be back-propagated from a time-domain loss.

We consider here two types of filterbanks: \emph{free} and \emph{analytic} \cite{pariente2020filterbank} ($\mathcal{A}$) along with the STFT.
In free filterbanks both analysis and synthesis parameters are unconstrained as in \cite{luo2018convtasnet} with $N$ fully learnable real and imaginary parts. For implementation purposes these are treated separately as $2N$ real filters and the whole filterbank is implemented as a 1D convolutional layer.

On the other hand, learnable analytic filterbanks have half of the filters fully learnable. For example, regarding the analysis filterbank $\left\{  \bm{\phi}_n(t) \right\}_{n=1, \hdots N}$, for each of the $N$ filters, the imaginary part is obtained from its real counterpart via the Hilbert transform $\mathcal{H}(\cdot)$:
\begin{equation}
    \bm{\phi}^{\mathcal{A}}_n(t) =  \bm{\phi}_n(t) + j  \mathcal{H}(\bm{\phi}_n(t)).
\end{equation}
The same is true for the synthesis filterbank $\left\{  \bm{\psi}_n(t) \right\}_{n=1, \hdots N}$. 
Because of this coupling, the modulus of a signal convolved with these learnt filters is invariant to small shifts in time domain, a property shared with the STFT. 
This property is crucial for the estimation of the SCMs in Equation \ref{eq:scm_estimation}, as the target signal mask is estimated on the reference channel and applied across all microphones.

The use of fully learnable filterbanks in place of the STFT poses some problems regarding the derivation of the SCMs. 
An implicit assumption for Eq. \ref{eq:scm_estimation} is that the analysis filterbank used is approximately orthogonal.
This condition is commonly referred to as the ``narrow-band approximation" and can be  satisfied by the STFT because of its approximate orthogonality \cite{kowalski2010beyond}.%under some assumptions \cite{}, mainly related to the maximum delay in the relative transfer function of the array and the length of the analysis window. 

Without this assumption, the target and interferer SCMs cannot be reduced to $M \times M$ matrices as in Eq.\ref{eq:scm_estimation},
as with no orthogonality of the basis, one must take into account also ``inter-frequency" terms. 
This leads to a block matrix SCM for each frame $k$ that can be partioned as an $N \times N$ block matrix (modeling the inter-frequency interactions) with the ($i$, $j$)-th block being a $M \times M$ matrix (modeling the inter-microphone interactions). 
This increases significantly the computational requirements as e.g. inversion of the full SCM leads to a complexity of $\mathcal{O}(N^3M^3)$ versus $\mathcal{O}(NM^3)$ for a diagonal block SCM.
%If the ``narrow-band approximation" holds, only the diagonal terms of such block matrix are significant and thus the SCM reduces to a diagonal block matrix for each frame, i.e. to an $M \times M \times N$ tensor as in Eq. \ref{eq:scm_estimation}.
%On the contrary, if this does not hold, the derivation of the MVDR and MWF beamforming solutions would be significantly more computationally expensive as e.g. inversion of the full SCM is required leading to a complexity of $\mathcal{O}(N^3M^3)$ versus $\mathcal{O}(NM^3)$ for a diagonal block SCM.

A straightforward, naive, but efficient approach, is to disregard the contribution of the ``inter-frequency" interactions in the SCM derivation also for the learned filterbanks. Since the filterbanks are learnt jointly with the rest of the model by minimizing a particular loss objective (e.g \ac{SISDR}) it can be assumed that the analysis filterbank will learn an approximately orthogonal basis. %i.e. for which the impact of the ``inter-frequency" interactions can be neglected. %and the full SCM reduces to be a block diagonal matrix.
We adopt in this work this rather strong assumption and provide some empirical evidence for this in Section \ref{sec:results}.

%A possible drawback of such learnable representations is that, contrary to the STFT, 
%This poses a problem as Equation \ref{eq:scm_estimation} is only valid as long as 

\vspace{-0.2cm}
\section{Experimental Setup}
\subsection{Datasets}
Crucially, most neural beamforming studies, being targeted mainly towards back-end tasks such as \ac{ASR}, perform their experiments using 16\,kHz signals. Such sampling rate however is sub-optimal for applications aimed towards human listening. %for which high frequency components have been shown to have perceptual significance \cite{}, impacting both the perceived quality of the speech signal as well as, to a lesser extent, intelligibility. 
For this reason, we use in our experiments the recently available First Clarity Challenge dataset \cite{akeroyd2020launching_clarity} which, being geared towards hearing aid development, is sampled higher at 44.1\,kHz.

%We argue however that such low sampling rate is hardly used in practical applications such as in laptops, smart-speakers et cetera, especially for applications where the enhanced signal is aimed towards human listening (e.g. hands-free communication) rather than automatic back-end tasks such as Automatic Speech Recognition (ASR) for which 16\,kHz are predominant. 
%Thus,  

%Such higher sampling frequency is also closed to practical applications such as in laptops, smartphones et cetera. 

% Samuele: I want to say that 44.1 and in general high freq is more used in actual devices e.g. laptops smarthphones et cetera and is also more challenging than 16kHz. 
\vspace{-0.2cm}
\subsubsection{Clarity Challenge Dataset}\label{sec:clarity_challenge_datase}
We use here the training and development subsets from the Clarity Challenge comprised of, respectively, 6k ($\sim10$ h) and 2.5k ($\sim4$ h) multi-channel simulated mixtures. 
Each simulated mixture consists in a target speaker and an interferer signal which can be either another competing speaker or a localized noise source. By dataset construction, each mixture is composed in such a way that the interferer signal always starts 2 seconds before the target signal. To make the task more challenging, in this work we only use 1 second of such ``preroll". 
Spatialization is performed using synthetic Room-Impulse-Responses (RIR) by simulating a randomized room with uniformingly sampled receiver, target and interferer locations, each constrained to be at least 1\,m apart from the others. The RIR RT60 has a log-normal distribution with a mean of 0.3 s and a standard deviation of 0.13 s. 
The Raven toolkit \cite{schroder2011raven} is used to perform such simulation. 
An array with a behind-the-ear hearing aid topology is employed with 3 microphones per ear. On each ear, microphones are spaced approximately $7.6$ mm (front, mid, rear) from one to another. 
We consider the task of recovering the reverberant target signal at one reference microphone without considering the head related impulse response. 
In this dataset, the SI-SDR at the array between the target and interferer signals has a -30 to 10\,dB range with a skewed gaussian distribution centered around 1\,dB. 
%Crucially, by dataset construction, in each example, 
In this work, we report results using the development and use a $90/10$ training set split for the purpose of training and validation respectively.

% short description of the corpora used
\subsection{Architecture and Training Details}
\label{sec:arcdetails}
 In our experiments we employ ConvTasnet \cite{luo2018convtasnet} separator as the mask-estimation \ac{DNN} in Figure \ref{fig:mask_beam}. 
 %The total number of parameters is $\sim 5.2$\,M. %As in \cite{luo2018convtasnet} we employ global layer normalization and standard layer normalization \cite{} for respectively non-causal and causal systems. 
 We train the whole system comprised of analysis, synthesis mask-estimation \ac{DNN} and beamforming solution in an end-to-end fashion using negated time-domain \ac{SISDR} \cite{le2019sdr} as the loss function. 
 Adam \cite{kingma2017adam} is used for optimization along with gradient clipping for gradients exceeding an $\mathcal{L}_2$ norm of $5$. We tune learning rate and weight decay for each experiment and train each model for a maximum of $100$ epochs with early stopping if no improvement is seen in the last $10$ epochs on the validation set. We halve the learning rate if no improvement is seen in the last $5$ epochs. 
 During training we randomly choose the reference channel from the $6$ available while 
 in testing and validation we always use the first left microphone as the reference.
 
\vspace{-0.2cm}
\section{Results}\label{sec:results}

In our experiments we consider, as an upper bound, MVDR and MWF beamformers with oracle \ac{WLM} in STFT domain. 
We use as performance metrics SI-SDR improvement (SI-SDRi) and Signal-to-Distortion Ratio \cite{vincent2006performance} improvement (SDRi). The SI-SDR and SDR values for no enhancement are respectively  $1.537$ dB and $1.144$ dB.

\begin{figure*}
\centering
\includegraphics[width=17cm]{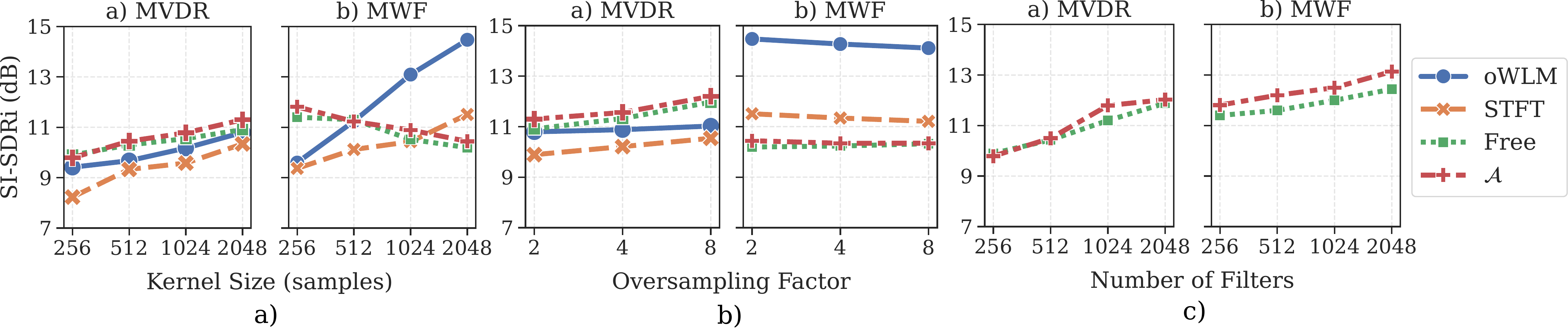}
\vspace{-0.2cm}
\caption{Performance for different MVDR and MWF configurations: oracle (oWLM) and learned models with different filterbanks (STFT, Free and $\mathcal{A}$). \\
\textit{a)} SI-SDRi versus kernel size. The number of filters is kept equal to kernel size and stride to half.
\textit{b)} SI-SDRi versus oversampling factor. The kernel size and number of filters is kept to 2048.
\textit{c)} SI-SDRi versus number of filters for learnable filterbanks. The kernel size and stride are kept fixed at 256 and 128 respectively.}%-based method here.}
%-based method here.}
\label{fig:sisdr_vs_overkernelnum}
\end{figure*}

In Figure \ref{fig:sisdr_vs_overkernelnum}a we report the SI-SDRi versus the length of the  analysis and synthesis filters (\emph{kernel size}) for different configurations. The number of filters is kept equal to the kernel size, and the stride half of that.

We can see that for both the STFT-based (\emph{STFT}) models and oracle (\emph{oWLM}) masks, performance improves as the kernel size increases. This is expected as a bigger kernel allows for more accurate SCMs estimation. 
%Interestingly the performance for MVDR STFT-based models is quite close to their oracle counterpart, but not for STFT-based MWF models: as the kernel size increases, the gap between the oracle and learned models widens. 
Both \emph{free} and $\mathcal{A}$ learned filterbanks outperform oracle WLM mask for small kernels. Only for MVDR, this is true also for all kernel sizes considered.
Interestingly, learned filterbanks seem to have opposing trends regarding MVDR and MWF in function of the kernel size. For MWF performance decreases as the kernel increases.
%This may be due to the fact that learning filterbanks with large kernel sizes is inherently more difficult and leads to more noisy training as far as MWF is considered. On the contrary, the MVDR distortion-less constraint can mitigate this issue. 

%The performance for \emph{free} configuration is especially surprising, as it has fully learned, decoupled real and imaginary parts and no shift-invariance inductive bias as the STFT or $\mathcal{A}$. 

%As expected it can be seen that the performance increases as the window length increases, as 

In Figure \ref{fig:sisdr_vs_overkernelnum}b we study how SI-SDRi changes by increasing the oversampling factor i.e. decreasing the stride while keeping fixed the kernel size. Here we fix the kernel size and number of filters to 2048 and vary the oversampling factor $N/H$ by 2, from 2 (same as in Figure \ref{fig:sisdr_vs_overkernelnum}a) to 8.

Regarding MVDR, for both STFT-based systems and \emph{oWLM} performance improves with higher oversampling but at a slower pace compared to what has been observed by increasing kernel size. Regarding MWF, performance decreases slightly for STFT and \emph{oWLM} while is almost constant for the models with learned filterbanks.

In Figure \ref{fig:sisdr_vs_overkernelnum}c we explore the effect of increasing the number of filters for learned filterbanks with fixed kernel size and stride of respectively 256 and 128 samples.
Such strategy is, in fact, one of the key factors which allows current monaural source separation algorithms to achieve such impressive performance \cite{luo2018convtasnet, luo2020dualpath}.

%Interestingly, theoretically this approach should have no advantages over using just as during optimization filters with  performed in \cite{} but we can speculate that learning filterbanks with large kernel sizes is inherently more difficult and leads to more noisy training. 

%In parallel, performance also improves to a certain extent by decreasing the stride at a fixed window size, i.e. increasing the STFT oversampling factor. However this improvement has a more modest effect and window length remains the most critical parameter. 

%In Table \ref{} we compare the best oracle configurations with performance 
%Figure \ref{} instead reports the same but for non-oracle systems: STFT-based ones 

For both beamforming solutions increasing the number of filters and thus forming an over-complete dictionary, improves significantly the performance. By comparing with Figure \ref{fig:sisdr_vs_overkernelnum}a, we can see that adding filters has a stronger effect with respect to expanding the kernel size. 
This suggests that beamforming with learned filterbanks may be particularly suited for low-latency applications as the kernel size can be kept low to suit the latency constraints, while the number of filters increased with no penalties in terms of latency. 

In Table \ref{tab:results_causal} we compare the best systems from previous experiments (Figure \ref{fig:sisdr_vs_overkernelnum}) in terms of both SI-SDRi and SDRi. As a term of comparison we also add iFasNet \cite{luo2020implicit}, a state-of-the art fully neural beamformer architecture. For this model we use the exact same configuration as in \cite{luo2020implicit}: as the sampling rate here is 44.1\,kHz here, iFasNet has more parameters compared to the original one due to increased window length. 

The proposed approach is competitive with the current state-of-the-art.
Among the non-oracle algorithms, MWF with learned filterbanks obtains the highest figures with the one based on analytic filterbank being the best. This latter consistently surpasses even the best oracle MVDR result.

\begin{table}
	\centering
	\setlength{\tabcolsep}{1.7pt} % Default value: 6pt
	\renewcommand{\arraystretch}{1} % Default value: 1
    \footnotesize
	\begin{tabular}{|ccccccc|}
		\hline
		Method  & N & L & H &  SI-SDRi [dB] &  SDRi [dB] & Params \\
		\hline
		\hline
	    oWLM-MVDR & 2048 & 2048  & 256 & 11.023  &  12.410 & - \\
	    oWLM-MWF & 2048 & 2048 & 1024 & 14.733 & 15.551 & -  \\
	    \hline
	    \hline
	   STFT-MVDR & 2048 & 2048  & 256 & 10.321    & 12.025 & 5.2M  \\
	   STFT-MWF & 2048 & 2048  & 1024 & 11.556 & 12.667  & 5.2M  \\
	    \hline
	   free-MVDR & 2048 & 256  & 128 & 11.882 & 12.963  & 6.3M \\
	   free-MWF & 2048 & 256  & 128 & 12.435 & 13.632  & 6.3M \\
	   \hline
	   $\mathcal{A}$-MVDR & 2048 & 256  & 128 & 12.024 & 13.372  & 5.8M  \\
	   $\mathcal{A}$-MWF & 2048 & 256  & 128  & \bf{13.142} & \bf{14.272} & 5.8M    \\
	    \hline
	    \hline
	    iFasNet \cite{luo2020implicit} & - & - & - & 9.896  & 10.342  & 4.4M \\
	    \hline
	\end{tabular}
	\vspace{-0.2cm}
	\caption{Comparison of best performing models in terms of SI-SDRi and SDRi and number of parameters (\emph{Params.}).}
	\label{tab:results_causal}

\end{table}

\begin{figure}
\centering
\includegraphics[width=7cm]{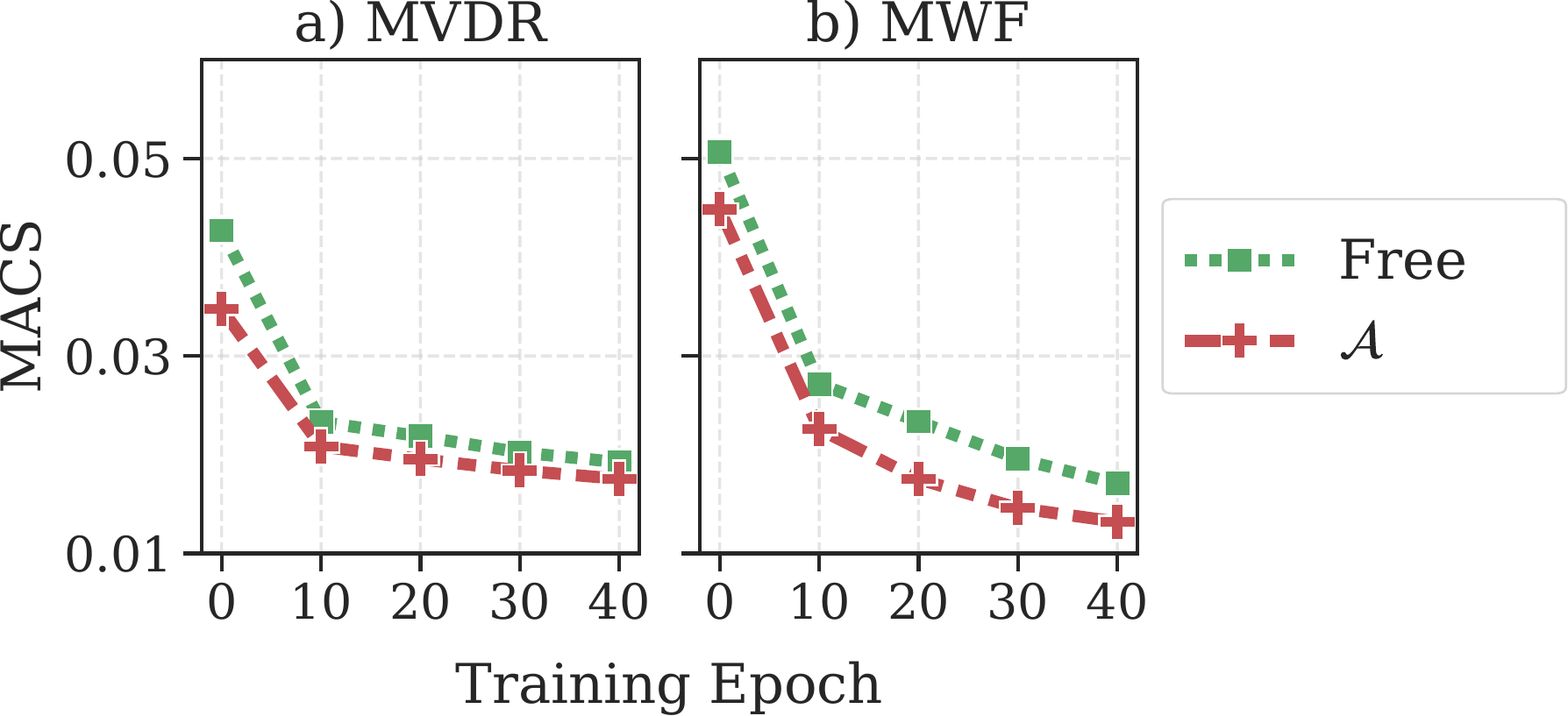}
\vspace{-0.2cm}
\caption{Mean Absolute Cosine Similarity (MACS) versus training epochs for learned filterbanks (Free and $\mathcal{A}$) MVDR and MWF models. All filterbanks have 1024 filters, 1024 kernel and 512 hop-size.}%-based method here.}
\vspace{-0.6cm}
\label{fig:MACS}
\end{figure}

\begin{figure}
\centering
\includegraphics[width=8cm]{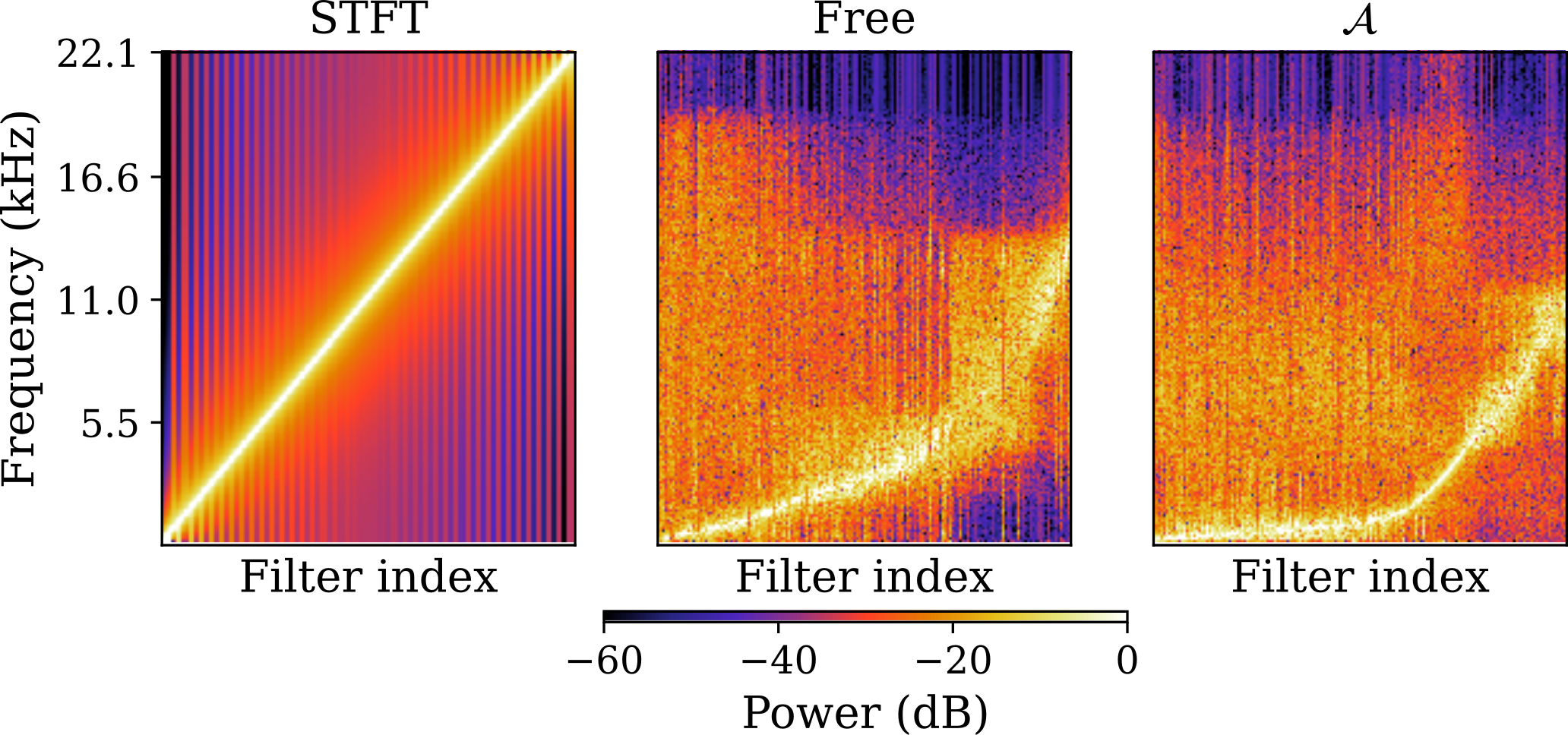}
\vspace{-0.2cm}
\caption{Frequency response of STFT, free and $\mathcal{A}$ filterbanks. All filterbanks have 2048 filters with 2048 samples kernel size.
For visualization purposes, filters in learned filterbanks are sorted according to their center-band frequency.}%-based method here.}
\label{fig:filterbank_response}
\vspace{-0.6cm}
\end{figure}

In Figure \ref{fig:MACS}, we report, at each training epoch, the Mean Absolute Cosine Similarity (MACS) for the analysis filterbanks of MVDR and MWF models with learned filterbanks. In detail, to measure the orthogonality of the learned filterbank, we compute the cosine similarity over each unique pair of filters, take the absolute value and take the average over the total number of unique pairs. 
We can see that the MACS value decreases during the training, indicating that the analysis filterbank gets more orthogonal as training progresses. This partly confirms the hypothesis made at the end of Section \ref{sec:leanrable}. On the other hand, the learned filterbanks converge, at best, to a MACS value of $0.013$ which is more than one order of magnitude higher than $0.001$, obtained for an STFT filterbank with same 1024 kernel size and number of filters. Future work could explore orthogonality constraints and their impact on performance.

Finally, in Figure \ref{fig:filterbank_response} we illustrate the frequency response of STFT and the learned filterbanks under study.
Both learned solutions tend to focus more on the lower part of the spectrum where most of speech energy is concentrated.
In fact, for free and $\mathcal{A}$, less filters are localized in the higher end of the spectrum, following loosely an exponential trend which is less steep than Mel-scale. This is especially true for $\mathcal{A}$ as most of the filters have a center-band frequency in the sub 2\,kHz region leading to an almost piece-wise linear trend. 
Free filters tend to have an higher frequency spread than analytic ones.

\section{Conclusions}

 In this work we proposed a fully end-to-end hybrid neural beamforming framework, where a DNN is employed to estimate the SCMs used to derive conventional beamforming solutions such as MVDR and MWF. 
 Differently from previous works, we consider the possibility to learn jointly with the DNN also the analysis and synthesis filterbanks instead of using the STFT and iSTFT. 
 %In particular, in this paper, we considered two types of learned filterbanks: fully learned ones, which don't have any constraint, and analytic ones, which, by design, display shift invariance. 
 We carried an extensive experimental study comparing learned filterbanks with STFT investigating how performance changes with different kernel sizes, stride factors and number of filters. 
 We found that such proposed strategy of performing spatial filtering in a learned representation is particularly effective for the MVDR beamformer. In fact, in this case we found learned filterbanks to consistently outperform STFT-based ones, even when oracle masks are employed. Regarding MWF, we found out that a gain over oracle masks is possible only for small kernel sizes. 
 This suggests that future work could explore causal, low-latency applications.
 Among the two learned filterbanks considered, the analytic ones fare the best. This promising result suggests that it may be worth exploring additional inductive biases for learned filterbanks such as orthogonality constraints. 
 
\section{Acknowledgements}
We would like to thank the Pytorch Audio team for useful discussion and for implementing native complex type support. 
%This work has been supported by the AGEVOLA project (SIME code 2019.0227), funded by the Fondazione CARITRO.

\bibliographystyle{IEEEbib}
\bibliography{refs}

\end{document}